# Characterization method to achieve simultaneous absolute PDE measurements of all pixels of an ASTRI Mini-Array camera tile

G. Bonanno[a*], G. Romeo[a], G. Occhipinti[a], M.C. Timpanaro[a], A. Grillo[a]

[a]*INAF, Osservatorio Astrofisico di Catania, Via S. Sofia 78, I-95123 Catania (Italy)*

**Abstract**

Recently, the Istituto Nazionale di Astrofisica (INAF) has placed a contract with Hamamatsu Photonics to acquire hundreds of Silicon Photomultipliers (SiPM) tiles to build 10 cameras with 37 tiles each for the ASTRI Mini-Array (MA) project. Each tile is made up of $8 \times 8$ pixels of $7 \times 7$ mm$^2$ with micro-cells of 75 μm. To check the quality of the delivered tiles a complex and acurate test plan has been studied. The possibility to simultaneously analyse as many pixels as possible becomes of crucial importance.

Dark Count Rate (DCR) versus over-voltage and versus temperature and Optical Cross Talk (OCT) versus over-voltage can be easily measured simultaneously for all pixels because they are carried out in dark conditions. On the contrary, simultaneous Photon Detection Efficiency (PDE) measurement of all pixels of a tile is not easily achievable and needs an appropriate optical set-up. Simultaneous measurements have the advantage of speeding up the entire procedure and enabling quick PDE comparison of all the tile pixels.

The paper describes the preliminary steps to guarantee an accurate absolute PDE measurement and the investigation the capability of the electronics to obtain simultaneous PDE measurements. It also demonstrates the possibility of using a calibrated SiPM as reference detector instead of a calibrated photodiode. The method to achieve accurate absolute PDE of four central pixels of a tile is also described.

## 1. Introduction

The ASTRI (Astrofisica con Specchi a Tecnologia Replicante Italiana) [3] Mini-Array (MA) is the next generation ground-based observatory for gamma-ray astronomy at very high energies. It will be based on nine telescopes, located in the northern hemisphere. ASTRI MA will build on the strengths of current Imaging Atmospheric Cherenkov Telescopes (IACT) and the accessible energy range is from few TeV to 300 TeV.

The nine telescopes will be installed at the Teide Astronomical Observatory, operated by the Instituto de Astrofisica de Canarias (IAC), on Mount Teide (~2400 m a.s.l.) in Tenerife (Canary Islands, Spain). The ASTRI MA will be the first array able to demonstrate the achievable performance of this series of telescopes when operated in synchronization. The array will give important results either in terms of scientific return or in terms of technological challenges. These telescopes have a relatively small plate scale that can be covered by SiPMs with sizes ranging from $6 \times 6$ mm$^2$ to $7 \times 7$ mm$^2$ [1].

SiPM detectors, thanks to their outstanding characteristics in terms of photon detection efficiency (PDE), photon number resolution, low operating voltage, fast dynamic response and insensitivity to magnetic fields, are suitable in the fields of high-energy astrophysics and IACT applications [2]-[6]. Considerable effort is presently being invested by the producers of SiPMs to further improve the performance achieved by this class of devices [7]-[9]; moreover, characterization studies and methodologies for evaluating the detector performance have been carried out [10]-[17]. Currently, we can safely assert that the SiPMs PDE is greater than that of Photomultiplier Tubes (PMTs) in the 300 ÷ 700 nm spectral range.

---

* *Corresponding author*: Giovanni Bonanno (giovanni.bonanno@inaf.it)

Recently, INAF has placed a contract with Hamamatsu Photonics to acquire 450 SiPM tiles to build 11 complete cameras made up of 37 tiles each for the above mentioned ASTRI MA project. Each tile consisting of $8 \times 8$ pixels of $7 \times 7$ mm$^2$ SiPM with micro-cells of 75 μm$^2$. A production of a huge amount of SiPMs is foreseen.

To check the quality of the delivered tiles a complex and accurate test plan has been studied. The possibility to simultaneously analyse as many pixels as possible becomes of crucial importance.

DCR versus over-voltage and versus temperature and OCT versus over-voltage can be easily measured simultaneously for all pixels because they are carried out in dark conditions; on the contrary, simultaneous PDE measurement of all pixels of a tile is not easily achievable and needs an appropriate optical set-up. Simultaneous measurements have the advantage of speeding up the entire procedure and enabling quick PDE comparison of all the tile pixels.

In this paper, we describe the preliminary steps to guarantee an accurate absolute PDE measurement, the investigation of the capability of the electronics to obtain simultaneous PDE measurements; we also show the possibility of using a calibrated SiPM as reference detector instead of a calibrated photodiode and finally, the method to achieve accurate absolute PDE of a certain amount of pixels located in the central area of a tile.

Three $3 \times 3$ mm$^2$ S14520 series Multi Pixel Photon Counting (MPPCs) have been tested and compared to prove the capability of the implemented set-up and procedure. One of these, the S14520-3050CS MPPC, has been selected as reference detector, and has been characterized by using a NIST calibrated photodiode; the other two MPPCs, the S14520-3050VN and the S14520-3075VN, have been simultaneously characterized using the reference MPPC to prove the correctness of the method.

Finally, we present the PDE measurements, at four wavelengths, carried out on a S14521-7075VN tile hosting only the central four uncoated MPPCs. The opto-mechanical apparatus designed on purpose and the complete procedure to achieve concurrent PDE measurements for all pixels have been also discussed. We demonstrated that the proposed technique can be applicable to SiPM arrays to speed up the PDE measurements. The adopted front end electronics allows us to acquire 32 simultaneous channels leading to a significant reduction in time and to a quick comparison among pixels.

## 2. ASTRI MA focal plane coverage and SiPM relevant mechanical parameters

The double-mirror optical configuration of the ASTRI MA telescopes allows an imaging resolving angular size of 0.19°. In order to match the angular resolution of the optical system, the design of the camera has to comply with the dimensions of the single pixel and of the basic detection module (the tile). Since the convex shaped focal surface of the camera has a curvature radius of about 1 m, the curved surface of the camera has to fit with a certain number of square pixels without losing the required focusing capability of the optical system. This specification is physically accomplished by a pixel of about $7 \times 7$ mm$^2$ and a detection module of $57.6 \times 57.6$ mm$^2$. A total of 37 detection modules (2368 pixels) form the camera at the focal plane; this structure is capable of achieving a full Field of View (FoV) of 10.5°. Using $7 \times 7$ mm$^2$ SiPM detectors with a pitch-size of 75 μm, 93 micro-cells fill completely a pixel sensitive area, giving a total of 8649 micro-cells. This ensures a sufficiently high dynamic range as required by the ASTRI MA project. An $8 \times 8$ tile will cover a total area of $57.6 \times 57.6$ mm$^2$ meaning 3317.76 mm$^2$, $6.975 \times 6.975$ mm$^2$ being the active area of each pixel, 0.2 mm the interspace between pixels and 0.2 mm the tile edge spacing. The total active area is 3091.36 mm$^2$ and thus a tile geometrical filling factor of 93.18 % is achieved. Figure 1 shows the tile schematics, a single $7 \times 7$ mm$^2$ pixel and the edge detail.

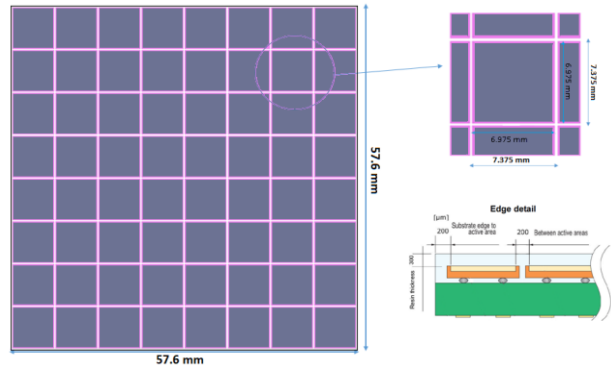

Figure 1. Tile schematics. Using 7 × 7 mm² SiPMs 57.6 × 57.6 mm² can be assembled. On the right part a detail of a single 7 × 7 mm² pixel is shown. With this configuration a total area of 3317.76 mm², an active area of 3091.36 mm² and a geometrical filling factor of 93.18 % are achieved.

## 3. SiPM detectors

The characterized detectors are of the same series, S14520, manufactured by Hamamatsu Photonics specifically optimized for Cherenkov light application.

Figure 2 shows the three 3 × 3 mm² SiPMs devices and a semi-populated tile hosting four 7 × 7 mm² MPPCs with 75µm microcells used to prove the suitability of the method. Apart from the MPPC S14520-3050CS, that is delivered in a ceramic package, all the other SiPMs are of the type Through Silicon Via (TSV). Table I reports the main characteristics of the detectors.

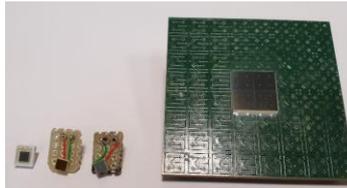

Figure 2. MPPCs of S14520 series devices: from left to right: 3050CS, 3050VN, 3075VN and the semi-populated tile with 4 7075VN pixels. Being the two 3x3 mm² non coated devices of the TSV type, we arranged a quick and simple assembly on a small bread board.

TABLE I
Main physical features of the characterized MPPC detectors

| device series | S14520-3050CS | S14520-3075VN | S14520-3050VN | S14521-7075AN |
|---|---|---|---|---|
| cell pitch | 50 μm | 75 μm | 50 μm | 75 μm |
| device size | 3 × 3 mm² | 3 × 3 mm² | 3 × 3 mm² | 7 × 7 mm² |
| number of pixels | 1 | 1 | 1 | 4 |
| sensitive area | 3.00 × 3.00 mm² | 3.00 × 3.00 mm² | 3.00 × 3.00 mm² | 4 x 6.975 × 6.975 mm² |
| micro-cells | 3600 | 3600 | 3600 | 8649/channel |
| contact type/package | wire bonding/ceramic | TSV | TSV | TSV Array |
| surface coating | Silicone resin | No coating | No coating | No coating |
| cell fill-factor | 74 % | 82 % | 74 % | 82 % each pixel |
| breakdown voltage* | 38.65 V | 38.94 V | 38.75 V | 38.48 V (average of 4) |

\* at 25°C

## 4. Characterisation Set-up

The electro-optical equipment used for SiPM measurements is depicted in Figure 3. Along the schematics, six main blocks can be envisaged: a) the SiPM and electronic front-end power supply, b) a Pulse generator, c) Laser and LED controllers, d) a calibrated reference detector and an electrometer to measure the photo-current, e) a CITIROC-1A electronic front-end to drive and process the SiPM signals, f) a counter to measure the dark count rate for the staircase. A set of calibrated neutral density filters are used in front of the SiPM to attenuate the signal and avoid working with a luminous level that could saturate the detector, while having sufficient current signal detected by the photodiode.

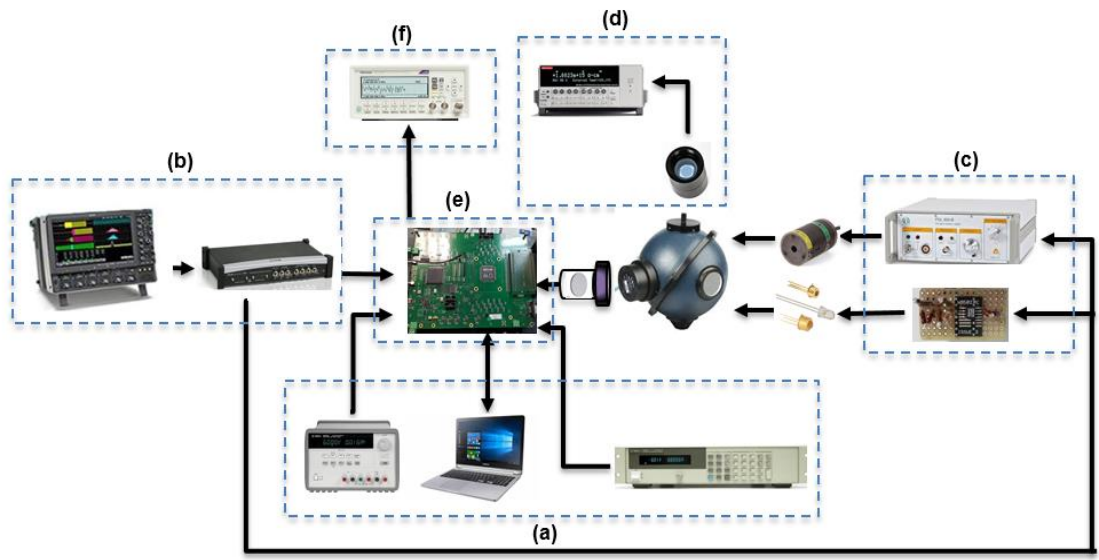

Figure 3. Characterization set-up based on pulsed light sources: lasers and LEDs. Apart from the integrating sphere, that can be considered the central part of the system, 6 main blocks can be envisaged: a) power supply, b) pulse generator, c) laser and LED drivers, d) calibrated reference photodiode, e) CITIROC 1A electronic front-end, f) counter.

The apparatus allows PDE measurements in the spectral range of 280 – 850nm by using 18 pulsed light sources (details are given in [1]).

As stated above, the main element of the front-end electronics for SiPM characterization is the Cherenkov Imaging Telescope Integrated Read-Out Chip (CITIROC) [17], produced by WEEROC[1]. The chip has been selected as the front-end electronics for the ASTRI camera. The analog core of the chip has 32 channels, each one incorporating: a digital-to-analog converter (DAC) for the SiPM high voltage adjustment, two preamps that allow setting dynamic range, through variable capacitors, between 16 fC and 320 pC, a trigger line consisting of two fast shapers (15 ns shaping time) and two discriminators, two slow shapers (from 12.5 ns to 87.5 ns shaping time). Two track and hold blocks are responsible for measurement of the Pulse Height Distribution (PHD).

### 5. Unavoidable preliminary step: photodiode re-calibration

Before using the above described equipment, we started with the first unavoidable step: the recalibration of the IRD photodiode through a NIST calibrated photodiode. This operation, normally, is done every 6 ÷ 10 months, but due to the carefulness of the measurements, we decided to start with the recalibration of the photodiode currently used.

The adopted set-up, depicted in the left part of Figure 4, is essentially based on an 8" integrating sphere illuminated by the pulsed sources with both photodiodes mounted to the two ports. In this case, two Keithley picoammeters are used to measure the current of the two photodiodes. The right part of Figure 4 reports the Quantum Efficiency (QE) versus wavelength in the 280 ÷ 850 nm spectral range of both photodiodes. The QE curve of the re-calibrated photodiode IRD 02#29 is used for all the SiPM PDE characterization.

---

[1] http://www.weeroc.com

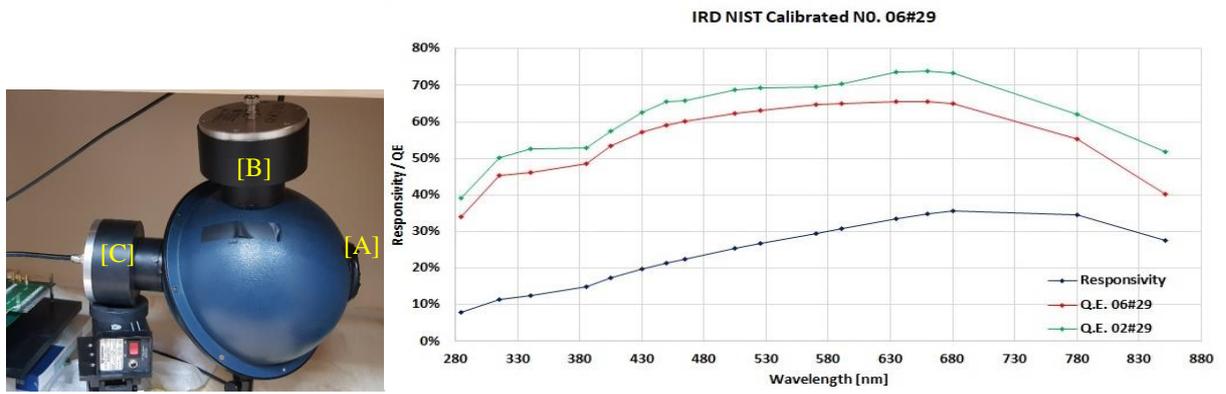

Figure 4. Left part: IRD UVG100 02#29 re-calibration through the NIST calibrated IRD UGV100 06#29 set-up: [A] input port for pulsed sources, [B] NIST Calibrated IRD photodiode 06#29, [C] IRD photodiode 02#29 to be recalibrated. Right part: QE versus wavelength in the 280 – 850 nm spectral range of both photodiodes, the NIST calibrated IRD 02#29 and the re-calibrated photodiode IRD 06#29

## 6. First attempt in simultaneous PDE measurements: $3 \times 3$ mm$^2$ MPPCs case

Before dealing with the MPPC array, we tried to check the method and the set-up with single $3 \times 3$ mm$^2$ SiPM detectors. We carried out simultaneous PDE measurements of two MPPCs, manufactured by using the same technology of those to be assembled in the tiles of the ASTRI MA cameras. The two MPPCs are uncoated and use the TSV technology. A third $3 \times 3$ mm$^2$ SiPM acting as reference detector, of the type S14520-3050CS, is also used. This last needs to be calibrated with the set-up shown in the left part of Figure 5. The calibrated photodiode is placed at one port of the integrating sphere while the mechanical housing hosting the SiPM is placed at the other port. A dedicated application software developed by WEEROC has been used to acquire the PHD of all 32 channels simultaneously. A detail of the mechanical housing hosting the device is depicted in the right part of Figure 5.

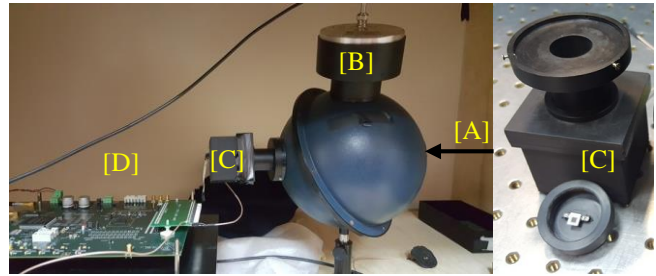

Figure 5. Set-up adopted to calibrate the S14520-3050CS MPPC: [A] input port for pulsed sources, [B] NIST Calibrated IRD photodiode 06#29, [C] S14520-3050CS MPPC, [D] CITIROC 1A evaluation board.

The measured PDE at 3.0 V of over-voltage in the 285 ÷ 850 nm spectral range is shown in Figure 8. The obtained PDE values will be considered as reference to calibrate other detectors of the same series.

SiPM absolute PDE measurements are based on the statistical analysis of the pulse heights distribution in light and dark conditions. In these measurements, the number of pulses per unit time in monochromatic light conditions are compared to the light level recorded at the same time, by a reference NIST photodetector. The used technique was discussed in detail in [1],[12],[18]. This technique is insensitive to the optical cross-talk, as we measure the probability of detecting 0 photons and derive the average assuming the Poisson statistics.

The average number of photons revealed by a SiPM is given by:

$$\bar{N}_{SiPM} = -ln\left[\frac{N_{ped}}{N_{ped}{}^{DK}}\right] \qquad (1)$$

Where: $N_{ped}$ is the number of events at 0 p.e. (i.e. the pedestal) during pulsed light measurements and $N_{ped}{}^{DK}$ is the number of events at 0 p.e. (i.e. the pedestal) without light source (dark).

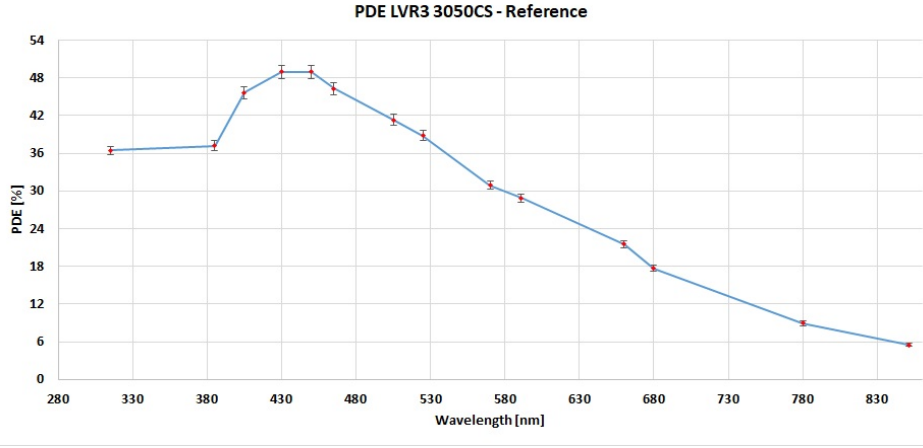

Figure 6. Absolute PDE of the S14520-3050CS MPPC at 3.0 V of over-voltage in the 285 ÷ 850 nm spectral range. This curve will be used as reference to obtain the PDE of other MPPCs of the same series.

The average number of photons revealed by calibrated photodiode $\bar{N}_{ph\,cal\text{-}Diode}$ is given by:

$$N_{ph\,Cal-Diode} = \frac{N_e}{QE(\lambda)} \qquad (2)$$

Where $N_e$ is the number of photoelectrons detected by the calibrated photodiode and $QE(\lambda)$ its Quantum Efficiency derived by the curve of Figure 4. The PDE is then obtained by the ratio:

$$PDE = \frac{\bar{N}_{SiPM}}{\bar{N}_{ph\,Cal-Diode}} \times \frac{A_{Diode}}{A_{SiPM}} \qquad (3)$$

Where $A_{Diode}$ and $A_{SiPM}$ are the areas of both detectors because we don't use diaphragms.

We can achieve the PDE of the other two SiPMs by simply carrying out a simultaneous measurement of both devices using the S14520-3050CS MPPC as reference detector. Note that in this case we don't use neutral density filter to attenuate the flux on the SiPM; so the uncertainty due to the transmittance measurement, is avoided.

We connected the reference MPPC to channel 0 and the S14520-3050VN to channel 1 of the CITIROC 1A board. The PDE measurements are obtained by simultaneously measuring the PHD of both channels. Figure 7 shows the PHD obtained for the two channels in dark conditions and by illuminating the integrating sphere with a 405 nm pulsed laser.

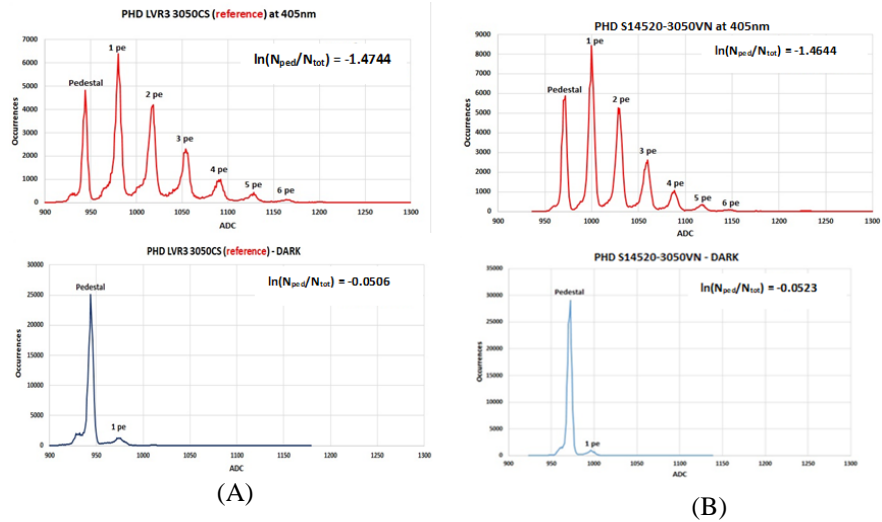

(A)                     (B)

Figure 7. PHD of both channels, 0 (panel A) and 1 (panel B), when illuminated with a pulsed source (upper distribution of both panels) and in dark conditions (lower distribution of both panels). These PHDs are obtained by using the WEEROC software application

By applying the formula (1) to the data derived from the PHD measurements (see Figure 10), we obtain the average number of photons $\bar{N}_{SiPM}$ for both channels, then the $PDE_{SiPM}$ of the detector under evaluation is given by the formula:

$$PDE_{SiPM} = \left[\frac{\bar{N}_{SiPM-ch1}}{\bar{N}_{SiPM-ch0}}\right] \times PDE_{REF} \qquad (4)$$

Where $\bar{N}_{SiPM-ch0}$ and $\bar{N}_{SiPM-ch1}$ are the average number of photons of $ch0$ and $ch1$, respectively.

In the specific case, $\bar{N}_{SiPM-ch0} = 1.42$ and $\bar{N}_{SiPM-ch1} = 1.41$ and being the $PDE_{REF}$ at 405 nm 45.60 % (see plot of Figure 6), from formula (4) we obtain $PDE_{SiPM} = 45.22$ %.

We made use of the above described procedure for each wavelength and derived the PDE in the 285–850 nm spectral range. We repeated the same measurements for the S14520-3075VN. The resulting PDEs are plotted in Figure 8. As expected the PDE of the S14520-3075VN is higher than that of the 50 μm microcell MPPC in the whole range due to the higher fill factor (see Table I).

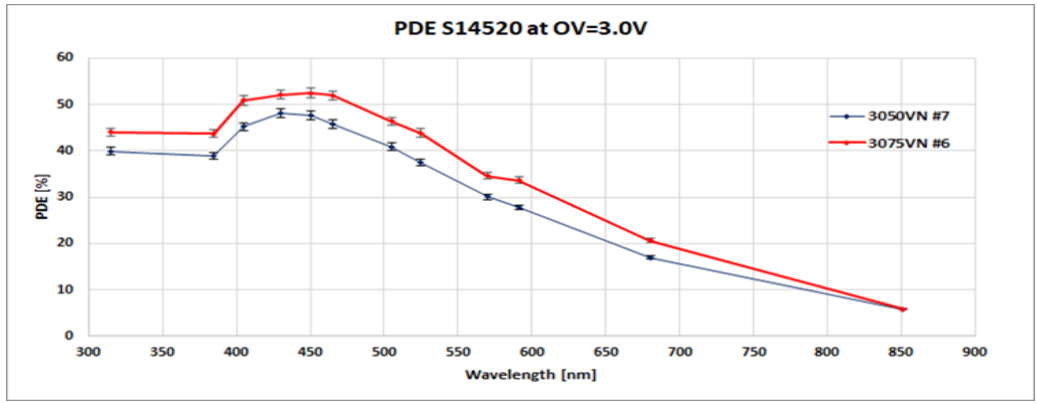

Figure 8. Absolute PDE in the 300 ÷ 850 nm spectral range of the two MPPCs S14520-3050VN#7 and S14520-3075VN#6, obtained using the S14520-3050CS as reference detector and using the method of simultaneous PHD measurements.

The achieved results lead to two quite relevant considerations: a) the method allows a more easily PDE characterization avoiding the use of neutral density filters and b) the device with 75 μm$^2$ microcell, non-coated and TSV, selected for the production of all the pixels of the ASTRI MA project, exhibits a PDE higher than 54% at 3.0 V of over-voltage in the 430 ÷ 470 nm.

### 7. Simultaneous PDE measurements of 7 × 7 mm$^2$ MPPCs assembled in a tile

To carry out concurrent PDE measurements of all the pixels assembled in a tile, we need to calibrate one pixel of the array by using the calibrated photodiode, then, following the above described procedure, we can measure the PDE of each remaining pixel at the same time. To prepare the set-up for this kind of measurement we designed a particular mechanical housing such that both detectors, the photodiode and the SiPM have the same aperture and optical path. Figure 9 shows the main parts of the apparatus: the semi-populated tile S14521-7075VN is placed at the center of the light-tight box and the top cover has a mechanical interface that allows the connection to the integrating sphere (left panel), the large aperture ensures the illumination of all the array without any vignetting (central panel). The SiPM board schematic (central panel) displays the number corresponding to the pixel position on the tile, the four central pixels are highlighted and pixel 37 has been selected as MPPC to be calibrated. In order to reproduce the same optical path of the reference photodiode, an eccentric adapter has been designed and realized to select a single pixel (right panel), that will be removed when concurrent PDE measurements are carried out.

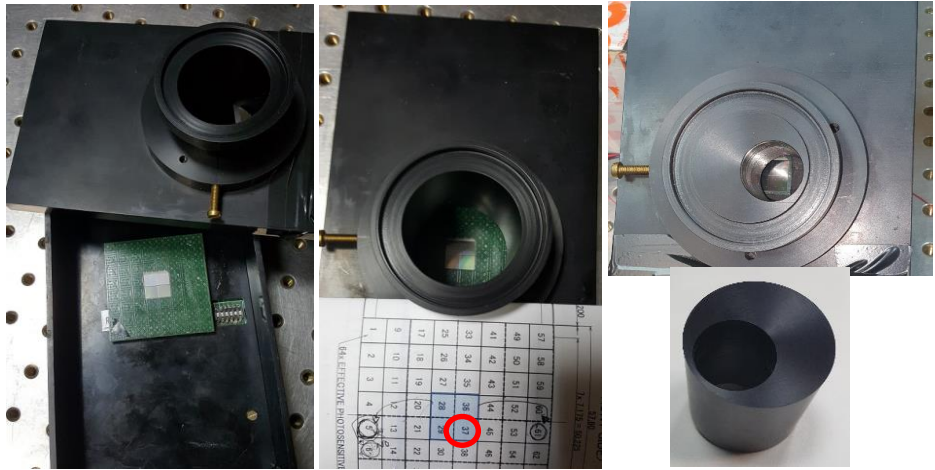

Figure 9. Set-up designed and realized to perform the simultaneous PDE measurements of the array S14521-7075VN MPPC.

The set-up to calibrate the lower right MPPC is depicted in the left panel of Figure 10, while the measured PDE is shown in the right panel. This PDE curve will be used as reference to measure the PDE of the other three pixels.

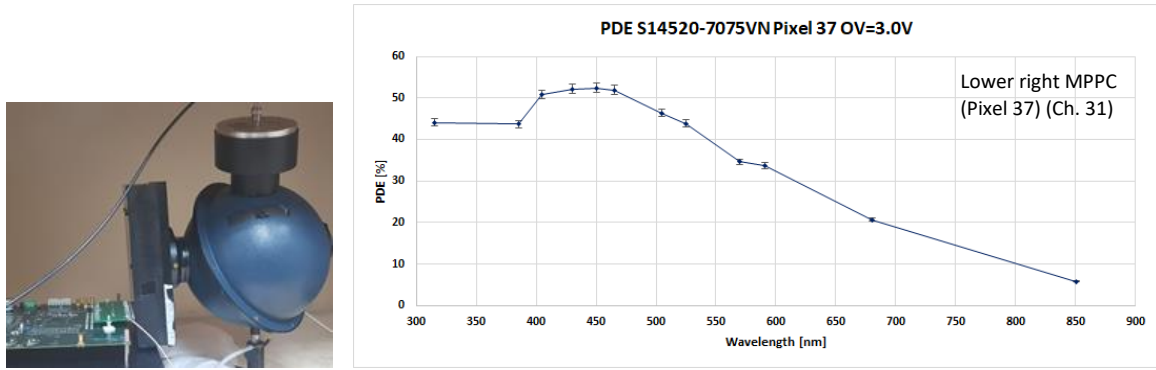

Figure 10. Left panel: set-up to calibrate the lower right pixel corresponding to the pixel No. 37. Right panel: absolute PDE of the tile's lower right MPPC at 3.0 V of over-voltage to be used as "reference" for the others MPPCs.

By removing the eccentric adapter, we can quickly perform the concurrent PDE measurement. The correspondence between pixel number (pix) and ASIC channel (ch) is as follows: pix 37 → ch 31, pix 29 →ch 1, pix 28 →ch 7, pix 36 →ch 25. The four PHDs are acquired simultaneously and, using the PDE curve of the lower right MPPC we measured the PDE. Only four pulsed sources have been utilized with wavelengths: 285 nm, 405 nm, 450 nm and 525 nm. The PDEs are plotted in Figure 11.

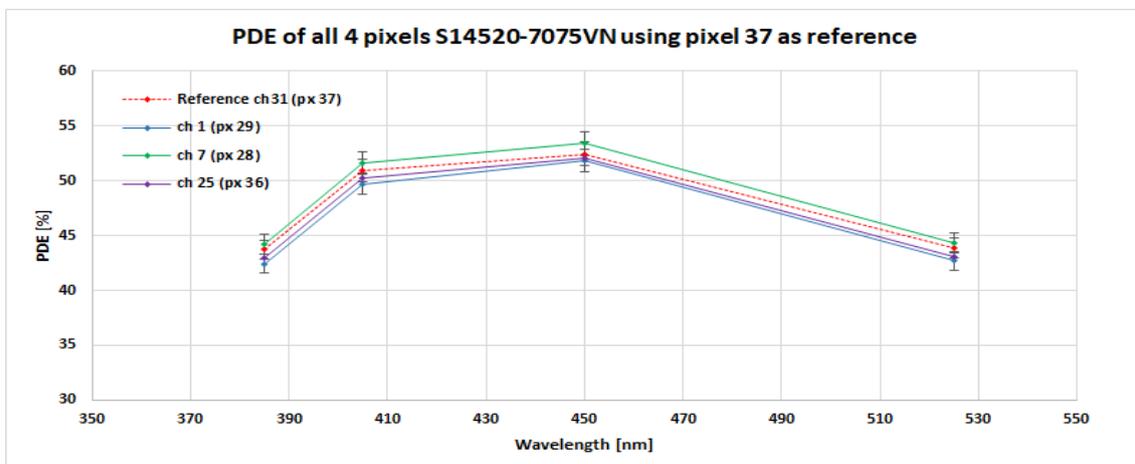

Figure 11. Absolute PDE of all four MPPCs at 3.0 V of over-voltage. The dotted line corresponds to the MPPC used as reference.

As it can be noted, and expected, a slight difference, within the experimental uncertainties, has been found between PDE values. The large aperture of the top cover box not introduces any vignetting up to 16 central pixels and thus, the procedure can give 16 simultaneous PDE values at each wavelength, which is extremely useful when a large amount of delivered tiles has to be checked.

## 8. Conclusions and Outlook

In this paper, measurement results of newly available large-area SiPM detectors are reported and discussed. The new Hamamatsu S14520 MPPC series achieve high photon detection efficiency and are suitable for IACT applications and in particular for dual-mirror Small-Sized Telescope like the ASTRI telescope.

The possibility to use a calibrated SiPM as reference detector instead of a calibrated photodiode has been discussed as well as the method to obtain concurrent PDE measurements. A more easily PDE characterization not affected by further uncertainty due to the neutral density filters is feasible. We proved that a PDE characterization of four central pixels of a tile is quite easy to achieve and that the apparatus is ready, without any change, to characterize up to sixteen central MPPCs. PDE measurements show that devices with a non-coated 75μm microcell, selected for the production of all the pixels of the ASTRI MA project, exhibit a PDE higher than 54% at 3.0 V of over-voltage in the 430 - 470 nm, promising very good scientific results.

**Acknowledgements**

This work is supported by the Italian Ministry of Education, University, and Research (MIUR) with funds specifically assigned to the Italian National Institute of Astrophysics (INAF) and by the Italian Ministry of Economic Development (MISE) within the "Astronomia Industriale" program. We acknowledge support from the Brazilian Funding Agency FAPESP (Grant 2013/10559-5) and from the South African Department of Science and Technology through Funding Agreement 0227/2014 for the South African Gamma-Ray Astronomy Programme.